\documentclass[usegraphicx,usenatbib,useapjfonts,apjl]{emulateapj}

\def\simlt{\mathrel{\rlap{\lower 3pt\hbox{$\sim$}}\raise 2.0pt\hbox{$<$}}}
\def\simgt{\mathrel{\rlap{\lower 3pt\hbox{$\sim$}} \raise
2.0pt\hbox{$>$}}}

\shorttitle{Primordial Non-Gaussianity and the NVSS}
\shortauthors{Xia {\it et al.}}

\begin{document}

\title{Primordial Non-Gaussianity and the NRAO VLA Sky Survey}

\author{Jun-Qing Xia\altaffilmark{1}, Matteo Viel\altaffilmark{2,3},
Carlo Baccigalupi\altaffilmark{1,2,3}, Gianfranco De
Zotti\altaffilmark{1,4}, Sabino Matarrese\altaffilmark{5,6} \& Licia
Verde\altaffilmark{7}}

\altaffiltext{1}{Scuola Internazionale Superiore di Studi Avanzati,
Via Bonomea 265, I-34136 Trieste, Italy; xia@sissa.it.}

\altaffiltext{2}{INAF-Osservatorio Astronomico di Trieste, Via G.B.
Tiepolo 11, I-34131 Trieste, Italy.}

\altaffiltext{3}{INFN/National Institute for Nuclear Physics, Via
Valerio 2, I-34127 Trieste, Italy.}

\altaffiltext{4}{INAF-Osservatorio Astronomico di Padova, Vicolo
dell'Osservatorio 5, I-35122 Padova, Italy.}

\altaffiltext{5}{Dipartimento di Fisica ``G. Galilei", Universit\`a
di Padova, Via Marzolo 8, I-35131 Padova, Italy.}

\altaffiltext{6}{INFN, Sezione di Padova, Via Marzolo 8, I-35131
Padova, Italy.}

\altaffiltext{7}{ICREA(Instituci\'o Catalana de Recerca i Estudis
Avan\c{c}ats) and Instituto de Ciencias del Cosmos, Universidad de
Barcelona, Marti i Franques 1, 08028, Barcelona, Spain.}

\begin{abstract}

The NRAO VLA Sky Survey (NVSS) is the only dataset that allows an
accurate determination of the auto-correlation function (ACF) on
angular scales of several degrees for active galactic nuclei
at $z \simeq 1$.  Surprisingly, the ACF is found to be positive on
large scales while, in the framework of the standard hierarchical
clustering scenario with Gaussian primordial perturbations it should
be negative for a redshift-independent effective halo mass of order
of that found for optically selected quasars. We show that a small
primordial non-Gaussianity can add sufficient power on very large
scales to account for the observed NVSS ACF. The best-fit value of
the parameter $f_{\rm NL}$, quantifying the amplitude of primordial
non-Gaussianity of local type, is $f_{\rm NL}=62 \pm 27$ ($1\,\sigma$
error bar) and $25<f_{\rm NL}<117$ ($2\,\sigma$ confidence level),
corresponding to a detection of non-Gaussianity significant at the
$\sim 3\,\sigma$ confidence level. The minimal halo mass of NVSS
sources is found to be $M_{\rm
min}=10^{12.47\pm0.26}h^{-1}M_{\odot}$ ($1\,\sigma$) strikingly
close to that of optically selected quasars. We discuss caveats and
possible physical and systematic effects that can impact on the
results.

\end{abstract}


%%%%%%%%%%%%%%%%%%%%%%%%%%%%%%%%%%%%%%%%%%%%%%%%%%

\section{Introduction}

The investigation of primordial non-Gaussianity offers a powerful
way of testing the generation mechanism of cosmological
perturbations in the early universe. Although the standard
single-field, slow-roll, canonical kinetic energy and adiabatic
vacuum state inflation generate very small non-Gaussianity, any
inflationary model that deviates from this may entail a larger level
of it \citep[and references therein]{bmr04,Komatsuwhitepaper}.

Deviations from Gaussian initial conditions are commonly taken to be
of the so-called local type and parameterized by the dimensionless
parameter $f_{\rm NL}$:
\begin{equation}
\Phi=\phi+f_{\rm NL}\left(\phi^2-\langle\phi^2\rangle\right)~,
\label{eq:fnl}
\end{equation}
where $\Phi$ denotes Bardeen's gauge-invariant potential and $\phi$
is a Gaussian random field. In this Letter, we use the cosmic microwave
background (CMB) convention for the quoted $f_{\rm NL}$ values.

%In the literature there are
%two conventions: in the large-scale structure (LSS) convention
%$\Phi$ is linearly extrapolated to $z=0$, while in the cosmic
%microwave background (CMB) convention it is evaluated deep in the
%matter era. Thus, $f^{\rm LSS}_{\rm NL}=[g(z=\infty)/g(z=0)]f^{\rm
%CMB}_{\rm NL}\sim1.3f^{\rm CMB}_{\rm NL}$, where $g(z)$ denotes the
%$\Lambda$-induced linear growth suppression factor.

A method \citep{DDHS08,MV08} for constraining non-Gaussianity from
large scale structure (LSS) surveys exploits the fact that the
clustering of extrema (i.e., dark matter halos where galaxies form)
on large scales increases (decreases) for positive (negative)
$f_{\rm NL}$. In particular, a non-Gaussianity described by Equation
(\ref{eq:fnl}) introduces a scale-dependent boost of the halo power
spectrum proportional to $1/k^2$ on large scales ($k<0.03\,h/$Mpc),
which evolves roughly as $(1+z)$.

Extragalactic radio sources are uniquely well suited to probe
clustering on the largest scales: (1) radio surveys are unaffected
by dust extinction which may introduce spurious features reflecting
the inhomogeneous extinction due to Galactic dust; (2) due to their
strong cosmological evolution, radio sources are very rare locally,
so that radio samples are free from the profusion of local objects
that dominate optically selected galaxy samples and tend to swamp
very large-scale structures at cosmological distances; (3) thanks to
the strong cosmological evolution, even shallow radio surveys reach
out to substantial redshifts. The NRAO VLA Sky Survey
\citep[NVSS;][]{Condon:1998iy} offers the most extensive sky
coverage (82\% of the sky to a completeness limit of about 3 mJy at
1.4 GHz) with sufficient statistics to allow an accurate
determination of the auto-correlation function (ACF), $w(\theta)$,
on scales of up to several degrees \citep{BlakeWall,Overzier}.

When a realistic redshift distribution of the NVSS sources is
adopted, the interpretation of the measured $w(\theta)$ in the
framework of the standard hierarchical clustering scenario with
Gaussian primordial perturbations requires an evolution of the bias
factor radically different from that of optically selected QSOs
\citep{Negrello,massardi}, in stark contrast with the similar
evolution of the luminosity function. In fact, the observed
$w(\theta)$ is positive up to large ($\sim 10^{\circ}$) angular
scales, which, for the median source redshift ($z_{\rm m}\simeq 1$),
correspond to linear scales where the correlation function should be
negative (see also \cite{hernandez-monteagudo}).
Here, we explore whether the (scale-dependent) large-scale
non-Gaussian halo bias could reproduce the observed shape of the
NVSS sources ACF, preserving the kinship with optically selected
active galactic nuclei (AGNs).

%Our work should be interpreted as a quantitative ``proof of
%principle'', highlighting the potential statistical power of the
%approach and its drawbacks and systematic uncertainties.

%%%%%%%%%%%%%%%%%%%%%%%%%%%%%%%%
\section{NVSS auto-correlation function}
%%%%%%%%%%%%%%%%%%%%%%%%%%%%%%%%%%%%%%%%

We include in our analysis only NVSS sources brighter than 10 mJy,
excluding the strip at $|b|<5^\circ$, where the catalog may be
substantially affected by Galactic emissions. This ensures a uniform
sky coverage \citep{BlakeWall}; the effect of possible residual
large-scale gradients producing and offset of the ACF is discussed in
\S\,\ref{sect:results}. The NVSS source surface density at this
threshold is $16.9\,{\rm deg}^{-2}$. The redshift distribution has
recently been determined by \citet{Brookes}.  Their sample, complete
to a flux density of 7.2 mJy, comprises 110 sources with $S_{1.4\rm
  GHz}\ge 10\,$mJy, of which 78 (71\%) have spectroscopic redshifts,
23 have redshift estimates via the $K$--$z$ relation for radio
sources, and 9 were not detected in the $K$ band and therefore have
only a lower limit in $z$. Here, we have adopted the description given
by \citet{DeZotti10}:
\begin{equation}
dN/dz=1.29+32.37z-32.89z^2+11.13z^3-1.25z^4~.
\end{equation}
The NVSS maps are pixelized using the HEALPix software package
\citep{healpix} with $N_{\rm side} = 64$, corresponding to $N_{\rm
  pix} = 49,152$ pixels with dimensions $0.92^\circ\times
0.92^\circ$.

The ACF estimator $\hat{w}(\theta)$ reads
\begin{equation}
\hat{w}(\theta)=\frac{1}{N_{\theta}}\sum_{i,j}\frac{(n_i-f_i\bar{n})(n_j-f_j\bar{n})}{\bar{n}^2} \;,
\end{equation}
where $f_i$ and $n_i$ are the coverage fraction and the number of
radio sources in each pixel, respectively; $\bar{n}$ is the
expectation value for the number of objects in the pixel
\citep[see][]{XiaISW}. The sum runs over all the pixels with a given
angular separation $\theta$. The equal weighting used here is
nearly optimal because of the uniform NVSS sky coverage and because
on large scales the noise is dominated by sample variance. For each
angular bin centered around $\theta$, $N_\theta$ is the number of
pixel pairs separated by an angle within the bin, weighted by the
coverage fractions.  We used $N_{\rm b} = 9$ angular bins in the
range $1^\circ\leq\theta\leq8^\circ$ with a linear binning plus
another estimate at $\theta=40'$, since below $40'$ the correlation
function is affected by multiple source components and above
$8^\circ$ the signal may be affected or even dominated by spurious
density gradients \citep{BlakeWall}.

\begin{figure}
\begin{center}
\includegraphics[scale=0.45]{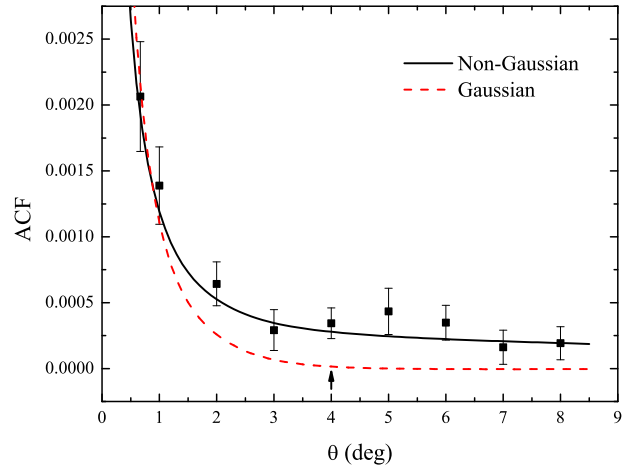}
\caption{Observed ACF of NVSS catalog. Values are jackknife
  estimated. The black solid line is the best fit model of our
  non-Gaussian calculations, while the red dashed line refers to the
  Gaussian case. The vertical arrow marks the angular scale above
  which the theoretical Gaussian ACF becomes negative. (Here, negative
  values are not visible, due to their very small amplitudes. See
  Figure \ref{fig:bias_one} for details.)}
  \label{fig:acf}
\end{center}
\end{figure}

We estimated the covariance matrix of the data points using the
jackknife re-sampling method \citep{covariance}. We divide the data
into $M=30$ patches, then create $M$ subsamples by neglecting each
patch in turn, and in each subsample we measure the ACF.  From the
$M$ estimates of the ACF functions, we estimate the diagonal
(variance) and off-diagonal (covariance) elements of the covariance
matrix for $\hat{w}(\theta)$.  In Figure \ref{fig:acf}, we plot the
observed NVSS ACF which is consistent with previous estimates
using different approaches \citep[e.g.][]{BlakeWall}. For
comparison, we also show the best fit theoretical ACF curve for the
Gaussian case ($f_{\rm NL}=0$) assuming the redshift-independent
minimal halo mass $M_{\rm min}$. Clearly, the curve does not match
the observed ACF data on large scales.

Note that the integral over the full survey solid angle covered by
the observationally determined ACF vanishes by construction. The
estimated values go to zero for $\theta\simeq 30^\circ$, and the
non-Gaussian model ACF shown in Figure \ref{fig:acf} becomes negative
approximately at the same $\theta$. However, no special meaning
should be attached to this coincidence since, as noted above, the
observational estimate of $w(\theta)$ are unreliable for $\theta
\simgt 8^\circ$.

%%%%%%%%%Figure%%%%%%%%%%%%%%%%%%
%%%%%%%%%Figure%%%%%%%%%%%%%%%%%%

\section{Method}
\subsection{Modeling the effects of non-Gaussianity}
The effects of non-Gaussianity on the source clustering properties
arise because a non-zero $f_{\rm NL}$ affects the halo mass function
and enhances the halo clustering on large scales. The second effect
is the dominant one.

In the presence of non-Gaussianity, the mass function $n_{\rm
  NG}(M,z,f_{\rm NL})$ can be written in terms of the Gaussian one
$n_{\rm G}^{\rm sim}(M,z)$, for which a good fit to the results of
simulations is provided by the Sheth-Tormen formula
\citep{shethtormen}, multiplied by a non-Gaussian correction factor
\citep{mvj,loverdeetal08}:
\begin{eqnarray}
&&\!\!\!\!\!\!R_{\rm NG}(M,z,f_{\rm NL})=1+\frac{\sigma^2_{\rm M}}{6\delta_{\rm ec}(z)} \cdot \nonumber\\
&&\!\!\!\!\!\!\cdot\!\left[\!S_{\rm
3,M}\!\left(\!\frac{\delta^4_{\rm ec}(z)}{\sigma^4_{\rm
M}}\!-\!2\frac{\delta^2_{\rm ec}(z)}{\sigma^2_{\rm
M}}\!-\!1\right)\! +\!\frac{dS_{\rm 3,M}}{d\ln{\sigma_{\rm
M}}}\!\left(\!\frac{\delta^2_{\rm ec}(z)}{\sigma^2_{\rm
M}}\!-1\!\right)\!\right]\!,
\end{eqnarray}
where the normalized skewness of the density field $S_{\rm
3,M}\propto f_{\rm NL}$, and $\sigma_{\rm M}$ denotes the rms of the
dark matter density field linearly extrapolated to $z=0$ and
smoothed on the scale $R$ corresponding to a Lagrangian radius of a
halo of mass $M$. Here, $\delta_{\rm ec}(z)$ denotes the critical
density for ellipsoidal collapse, which for high peaks is
$\delta_{\rm ec}(z)\sim\delta_{\rm c}(z)\sqrt{q}$ ($q=0.75$) and has
been calibrated on N-body simulations \citep{grossi09} and
$\delta_{\rm c}(z)=\Delta_{\rm c}(z)D(0)/D(z)$ where $D(z)$ denotes
the linear growth factor; $\Delta_{\rm c}(z)\sim 1.68$ and evolves
very weakly with redshift.

More importantly, the large-scale halo bias is also modified by the
presence of non-Gaussianity \citep{DDHS08,MV08,grossi09}:
\begin{equation}
b_{\rm NG}(z)-b_{\rm G}(z)\simeq2(b_{\rm G}(z)-1)f_{\rm
NL}\delta_{\rm ec}(z)\alpha_{\rm M}(k)~, \label{eq:nghalobias}
\end{equation}
where the factor $\alpha_{\rm M}(k)$ encloses the scale and halo
mass dependence. In practice, we find that, on large scales,
$\alpha_{\rm M}(k)\propto 1/k^2$ and is independent of the halo
mass.

%%%%%%%%%Figure%%%%%%%%%%%%%%%%%%
\begin{figure}
\begin{center}
\includegraphics[scale=0.45]{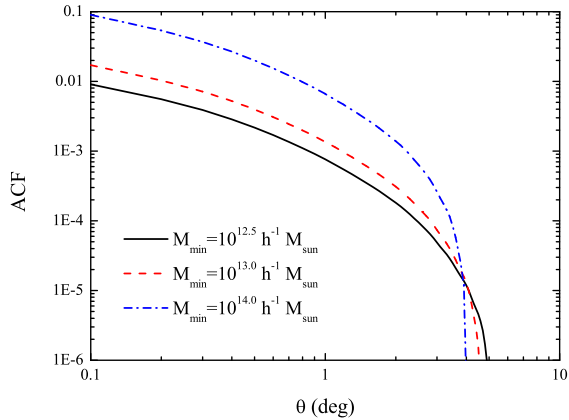}
\caption{Effect of different M$_{\rm min}$ masses on the ACF: the
zero-crossing angular scale of the ACF decreases with increasing
$M_{\rm min}$ for $f_{\rm NL}=0$.} \label{fig:bias_one}
\end{center}
\end{figure}

We start by assuming that the large-scale, linear halo bias for the
Gaussian case is \citep{shethtormen}
\begin{eqnarray}\label{eq:bST}
b_{\rm G}&=&1+\frac{1}{D(z_{\rm o})}\left[\frac{q\delta_{\rm c}(z_{\rm f})}{\sigma^2_{\rm M}}-\frac{1}{\delta_{\rm c}(z_{\rm f})}\right]\nonumber\\
&&+\frac{2p}{\delta_{\rm c}(z_{\rm f})D(z_{\rm o})}\left\{1+\left[\frac{q\delta^2_{\rm c}(z_{\rm f})}{\sigma^2_{\rm M}}\right]^p\right\}^{-1}~,
\end{eqnarray}
where $z_{\rm f}$ is the halo formation redshift, and $z_{\rm o}$ is
the halo observation redshift. As we are interested in massive
halos, we expect that $z_{\rm f}\simeq z_{\rm o}$. Here, $q=0.75$
and $p=0.3$ account for non-spherical collapse and are a fit to
numerical simulations \citep[see also][]{Mo96,Mo97,Scoccimarro01}.
We will later relax this assumption.

Finally, the weighted effective halo bias is given by
\begin{equation}
b_{\rm NG}^{\rm eff}(M_{\rm min},z,k,f_{\rm
NL})=\frac{\int^\infty_{M_{\rm min}}b_{\rm NG}n_{\rm
NG}dM}{\int^\infty_{M_{\rm min}}n_{\rm NG}dM}~.
\end{equation}

Two things should be clear from Equation (\ref{eq:nghalobias}): (1) there is
a degeneracy between $b_{\rm G}$ and $f_{\rm NL}$ (the same amount of
non-Gaussian bias can be given by different pairs of $b_{\rm G}$,
$f_{\rm NL}$ values; strictly speaking, $b_{\rm G}$ is not a free
parameter here, and the degeneracy is between $M_{\rm min}$, which is
a free parameter, and $f_{\rm NL}$; however $b_{\rm G}$ is strongly
dependent on $M_{\rm min}$); (2) the $1/k^2$ scale dependence means
that large-scales are mostly affected by $f_{\rm NL}$ and small scales
are primarily affected by $b_{\rm G}$. A positive $f_{\rm NL}$
enhances the amplitude of auto-correlation power spectra especially at
large angular scales ($\ell<200,~\theta>4^\circ$). This is the effect
we shall use to constrain $f_{\rm NL}$ and its impact on the ACF is
clearly visible in Figure \ref{fig:acf}. In fact, in the Gaussian case,
for the adopted redshift distribution and a redshift-independent
$M_{\rm min}$, the ACF is expected to become negative for
$\theta>4^\circ$.  This is also shown in Figure \ref{fig:bias_one} where
the ACF is plotted for several values of $M_{\rm min}$.

%%%%%%%%%Figure%%%%%%%%%%%%%%%%%%
\begin{figure}
\begin{center}
\includegraphics[scale=0.45]{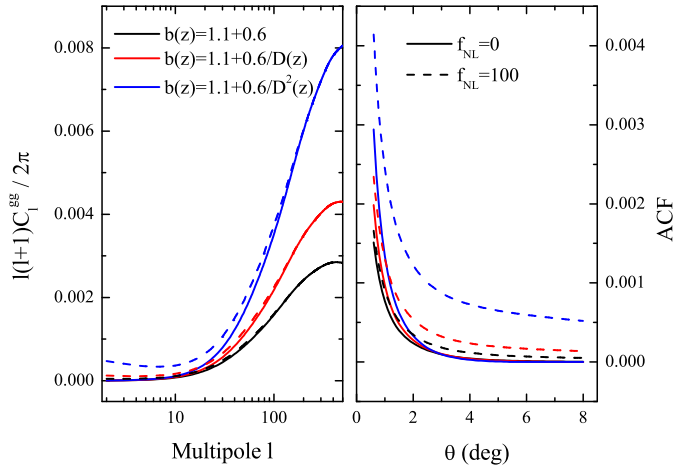}
\caption{Effects of non-Gaussianity on the auto-correlation power spectra (left
panel) and on the ACF (right panel) for three different Gaussian
bias models.} \label{fig:bias_two}
\end{center}
\end{figure}
%%%%%%%%%Figure%%%%%%%%%%%%%%%%%%

In general, the expected degeneracy between $f_{\rm NL}$ and $b_{\rm
G}$ may be lifted in two ways:  (1) on small scales the effect of
$f_{\rm NL}$ is completely negligible but not that of $b_{\rm G}$,
(2) the redshift dependences of the two contributions are different.
However, the angular correlation function encompasses the signal from
different redshifts and different physical scales, complicating the
separation of the $f_{\rm NL}$ and $b_{\rm G}$ signals.

To explore the effect of relaxing the assumption $z_{\rm f} \simeq
z_{\rm o}$, we have also considered a model for the Gaussian bias given
by \citep[and references therein]{MCLM1,MCLM2}
\begin{equation}
b_{\rm G}(z)=b_1+\frac{b_2}{D^{\gamma}(z)}
\,\,\,\,\,\,(0\leq\gamma\leq2), \label{eq:biasgeneral}
\end{equation}
with $b_1$ and $b_2$ being free parameters. Indeed, an
``object-conserving" bias model corresponds to $\gamma \approx 1$,
while the bias of high-density peaks for objects that have just
formed yields $\gamma \approx 2$. In Figure \ref{fig:bias_two}, we show
the effect of a Gaussian bias model on the auto-correlation power
spectra and ACF when varying the power-law index $\gamma$; the
larger the value of $\gamma$ the larger is the large-scale
non-Gaussian boost.

\citet{lin01} and \citet{yoo09} discussed the gauge dependence of
matter power-spectrum on very large scales ($k<0.003\, h/{\rm
  Mpc}$). We find that this gauge-dependent effect on the matter
power spectrum can be mimicked by that of a non-Gaussian halo bias
model with $f_{\rm NL}\sim5$. Here, we calculate the matter
power spectrum in the conformal Newtonian gauge.

%%%%%%%%%%%%%%%%%%%%%%%%%%%%%%%%%%%%%%%%%%%%%%%%%%%%%%%%%%%%%%%%%%%
\subsection{Implementation and data sets}
\label{sec:results} The theoretical prediction for  the ACF depends
on: the  cosmological parameters, the minimal halo mass $M_{\rm
min}$ and $f_{\rm NL}$. For the generalized bias model of
Equation (\ref{eq:biasgeneral}), we also add the $b_1$ and $b_2$ and
$\gamma$ bias parameters.

Rather than fixing the cosmological parameters to the best-fit
values derived from Wilkinson Microwave Anisotropy Probe seven year (WMAP7), we will report the
results after having marginalized over them with a prior given by a
compilation of recent data sets.

We perform a global fitting using the {\tt CosmoMC} package
\citep{cosmomc}, a Markov Chain Monte Carlo code, which has
been modified to calculate the theoretical ACF. We assume purely
adiabatic initial conditions and a flat universe, with no tensor
contribution. We vary the following cosmological parameters
$(\Omega_{\rm b} h^2,\,\Omega_{\rm c} h^2,\,\tau,\,\Theta_{\rm
s},\,n_{\rm s},\, A_{\rm s},\, f_{\rm NL},\, M_{\rm min})$, where
$\Omega_{\rm b} h^2$ and $\Omega_{\rm c} h^2$ are the baryon and
cold dark matter densities, $\tau$ is the optical depth to
reionization, $\Theta_{\rm s}$ is the ratio (multiplied by 100) of
the sound horizon at decoupling to the angular diameter distance to
the last scattering surface, $n_{\rm s}$ is the primordial spectral
index, and $A_{\rm s}$ is the primordial amplitude. We do not consider
massive neutrinos and dynamical dark energy and for the pivot scale
we set $k_{\rm s0}=0.05\,$Mpc$^{-1}$.

We also use (1) CMB temperature and polarization angular power
spectra as measured by WMAP7 \citep{WMAP7}, (2) baryonic acoustic
oscillations in the galaxy power spectra as measured by the SDSS7
and the Two-degree Field Galaxy Redshift Survey
\citep[2dFGRS;][]{BAO} (3) SNIa distance moduli of Union compilation
from the Supernova Cosmology Project \citep{Kowalski:2008ez}. We add
a prior on the Hubble constant, $H_0=74.2 \pm 3.6$ km/s/Mpc
\citep{HST}. Finally, we set the minimal halo mass $M_{\rm min} >
10^{12} h^{-1}M_{\odot}$ consistent with observations showing that
radio AGNs are hosted by halos more massive than those hosting
optical QSOs \citep{Hickox}.

%%%%%%%%%%%%%%%%%%%%%%%%%%%%%%%%%%%%%%%%%%%%%%%%%%%%%%%%%%%%%%%%%%%%%%%%%%%%
\section{Results}\label{sect:results}
%%%%%%%%%Figure%%%%%%%%%%%%%%%%%%
\begin{figure}
\begin{center}
\includegraphics[scale=0.4]{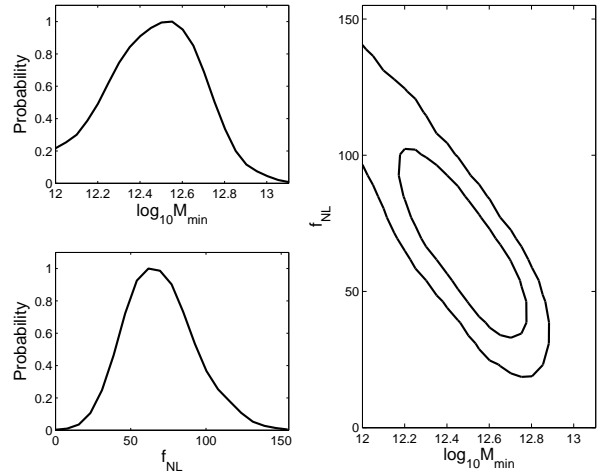}
\caption{Marginalized one-dimensional and two-dimensional
  distributions (1, 2$\,\sigma$ contours) of the minimal halo mass
  $M_{\rm min}$ and of the non-Gaussian parameter $f_{\rm
    NL}$. \label{fig:result}}
\end{center}
\end{figure}
%%%%%%%%%Figure%%%%%%%%%%%%%%%%%%

We start by considering the case where the Gaussian bias is given by
Equation (\ref{eq:bST}). In Figure \ref{fig:result}, we show the
one-dimensional posterior probability distributions for $M_{\rm
min}$ and $f_{\rm NL}$ after marginalizing over the other
parameters. The right panel shows the degeneracy between $M_{\rm
min}$ and $f_{\rm NL}$. Note that the constraints on $f_{\rm NL}$
only come from the ACF data; external data sets are only used to set
the underlying cosmology.

We find that the current ACF implies $f_{\rm NL} > 0$ at the $\sim
3\,\sigma$ confidence level. The reason for that can be clearly seen
in Figure \ref{fig:bias_one} where the Gaussian model with the
redshift-independent $M_{\rm min}$ yields an ACF dropping to zero
for $\theta \simeq 4^\circ$ and becoming negative on larger scales
where the observed ACF is still positive. Non-Gaussianity of the
local type adds power on large angular scales yielding a good fit to
the observed data points.

The marginalized constraints on the non-Gaussianity parameter
$f_{\rm NL}$,
\begin{equation}
f_{\rm NL}=62\pm27~(1\,\sigma~~{\rm CL})~,
\end{equation}
\begin{equation}
(6)25<f_{\rm NL}<117(142)~~[95\%~(99.7\%)~~{\rm CL}],
\end{equation}
are compatible with other previous estimates
\citep{Yadav:2007yy,Slosar,Curto:2009pv,Smidt:2009ir,JV,Smith:2009jr,Rud}
and in very good agreement with the recent WMAP7 estimate
\citep{WMAP7}.

The minimal effective halo mass, $M_{\rm
  min}=10^{12.47\pm0.26}h^{-1}M_{\odot}$ ($1\,\sigma$), turns
out to be remarkably close to that found for optically selected QSOs:
$M_{\rm QSO}=(3.0\pm 1.6)\times 10^{12}h^{-1}M_{\odot}$
\citep{Croom}.

To explore whether a more general bias model (with $\gamma$ allowed
to conservatively vary even in an un-physical range $\gamma<1$) than
that of Equation (\ref{eq:bST}) may reconcile a Gaussian model with
the data, we repeated the analysis using the bias of Equation
(\ref{eq:biasgeneral}). We keep $b_1$ and $b_2$ fixed to 1.1 and
0.6, respectively, and we vary $0<\gamma<2$ (the values for $b_1$
and $b_2$ have been chosen in order to provide a good fit to the
data). In this case, the recovered central value of $f_{\rm NL}$
becomes a little higher and the $1\,\sigma$ error bar increases by
about a factor of 2 (a smaller $f_{\rm NL}$ accommodates larger
$\gamma$ and larger bias).

We also perform a cross-check by using the published version of the
ACF of \citet{BlakeWall} (in this case zero covariance between data
points is assumed, since no covariance matrix has been computed) and
find $f_{\rm NL}=58 \pm 12\, (1\,\sigma\,$ CL) (we rely on 12 data
points in the range $0.5623^\circ \leq \theta \leq 7.70795^\circ$).
If we instead set the off-diagonal terms of our ACF estimate to
zero, we get $f_{\rm NL}=70 \pm 15 \, (1\,\sigma\,$ CL). From these
results we can conclude that covariance between data points
increases the error bar on $f_{\rm NL}$ by almost a factor of 2 and
the \citet{BlakeWall} and our ACF measurements are in very good
agreement with each other in terms of derived $f_{\rm NL}$ values.
Another instructive cross-check is to see to what extent our
conclusions are affected by applying an overall subtraction to all
the ACF values of $10^{-4}$, in order to correct for a possible
systematic offset that can contaminate the signal \citep{BlakeWall}.
In this case, our constraints are weaker but consistent with the
previous analysis and we get $f_{\rm NL}=42 \pm 30 \, (1\,\sigma\,$
CL). We have also checked that the correction proposed by
\citet{ws09} to account for the infrared divergence of the
non-Gaussian halo correlation function is negligibly small for our
best-fit $f_{\rm NL}$ value.

To allow for the ``integral constraint'' (measurements probe the
survey mean, and not the ensemble mean), we have added to
$w(\theta)$ a constant $c$ and have marginalized over $c$, allowing
this quantity to vary in the range $[10^{-8},10^{-4}]$ (the upper
limit cannot be larger since this is the theoretical variance
expected on the scale of the survey scales). We find $f_{\rm
NL}=58\pm28$ ($1\,\sigma$), showing that, as expected given the
large sky fraction covered by the NVSS survey, the best-fit value of
$f_{\rm NL}$ is only marginally affected.

%%%%%%%%%%%%%%%%%%%%%%%%%%%%%%%%%%%%%%%%%%%%%%

We use a Fisher matrix approach to estimate which scales contribute
most to the signal for $f_{\rm NL}$ and $\log_{10}M_{\rm min}$ from
the ACF as a function of $\theta$. Depending on where the signal is
localized, this may give some insights into what systematic effects
are important.

\begin{figure}
\begin{center}
\includegraphics[scale=0.45]{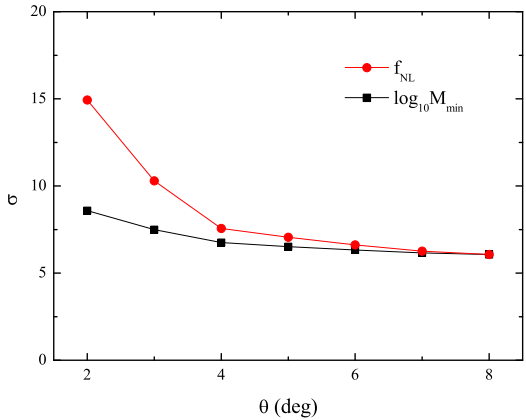}
\caption{Dependence on the maximum separation $\theta$ of the error
bars on the minimal halo mass $M_{\rm min}$ and on $f_{\rm NL}$,
calculated with the Fisher matrix method (arbitrary normalization).}
\label{fig:error}
\end{center}
\end{figure}

Figure \ref{fig:error} shows the Fisher-predicted error (for the ACF
of a survey with NVSS characteristics) as a function of the maximum
angle $\theta$ (the minimum is always set to $1^{\circ}$). The
error-bar normalization is arbitrary and error bars of $f_{\rm NL}$
and $\log_{10}M_{\rm min}$ at $\theta=8^\circ$ are set to be the
same. The error on $M_{\rm min}$ stabilizes at $\theta \sim
3^{\circ}$, indicating that the bias signal is mainly localized at
separations smaller than $3^{\circ}$. The $f_{\rm
  NL}$ error decreases  rapidly and stabilizes at larger $\theta$,
indicating that the non-Gaussian signal is localized in the ACF at
$\theta=2^\circ-5^\circ$.  Since the mean redshift of NVSS sources
brighter than 10 mJy is about $1.2$, $1^\circ$ corresponds to a
comoving size $r\sim 60\, {\rm Mpc}$. The maximum non-Gaussianity
signal thus comes from comoving scales in the range $120\,{\rm
Mpc}<r<300\,{\rm Mpc}$, while the constraints on $M_{\rm min}$ come
primarily from scales $r<180\, {\rm Mpc}$.

%%%%%%%%%Figure%%%%%%%%%%%%%%%%%%
%%%%%%%%%Figure%%%%%%%%%%%%%%%%%%

\section{Discussion and Conclusions}

While previous analyses exploiting the NVSS to constrain primordial
non-Gaussianity of local type have focused on the cross-correlation
with the WMAP Internal Linear Combination map, we have shown that
the angular correlation function alone is a very sensitive
non-Gaussianity probe. The key point is that, given the redshift
distribution of NVSS, which has been recently observationally
determined, and the typical AGN halo mass estimated for optically
selected QSOs, the standard $\Lambda$CDM cosmology with
scale-independent bias would imply, on scales $>4^\circ$, a negative
ACF, but, on the contrary, it is observed to be positive. Careful
analyses of the NVSS sample \citep[e.g.][]{BlakeWall} indicate that
systematic offsets that may induce a spurious positive signal should
be negligible for the sources with $S_{1.4\rm GHz}>10\,$mJy,
selected for the present analysis. If so, the NVSS ACF may point at
the presence of a small primordial non-Gaussianity that adds power
to the largest scales. After marginalizing over all the other
parameters, we find $25 < f_{\rm NL} < 117$ at the $95\%$ confidence
level, compatible with bounds derived by other studies. The minimum
halo mass turns out to be $M_{\rm min}=10^{12.47\pm
0.26}\,h^{-1}\,M_\odot$ ($1\,\sigma$), remarkably close to the value
found by \citet{Croom} for optically selected QSOs.

We have addressed the significance and the robustness of our
findings by considering different bias models and by investigating
the impact of gauge effects on large scales.  Error bars were
estimated by a jackknife re-sampling procedure, widely used in the
literature \citep{Scrantonetal03,XiaISW}. It is known to be robust
and accurate for the diagonal elements of the covariance matrix but
not as widely tested and calibrated for the off-diagonal ones, which
cannot be neglected because neighboring ACF data points are highly
correlated. From \cite{Scrantonetal03}, we infer that jackknife can
underestimate parameter errors by up to 30\%. If the error bars were
to be increased by this (maximal) amount, $f_{\rm NL}$ would become
compatible with zero at the $\sim2\,\sigma$ confidence level. We
conclude that our work should be seen as a ``proof of principle",
indicating that future surveys probing scales $\sim 100$ Mpc at
substantial redshifts can put stringent constraints on primordial
non-Gaussianity \citep[e.g.][]{carbone08,viel09}.

%%%%%%%%%%%%%%%%%%%%%%%%%%%%%%%%%%%%%%%%%%%%%%%%%%%%%%%%%%%%%%%%%%%%%%%%%%%%

{\it Acknowledgments: } We thank the referee for constructive
comments. We acknowledge the use of LAMBDA and the HEALPix package.
Numerical analysis has been performed at the University of Cambridge
High Performance Computing Service. Research is supported by the ASI
Contract No. I/016/07/0 COFIS, the ASI/INAF Agreement I/072/09/0 for
the Planck LFI, PRIN MIUR, MICCIN grant AYA2008-0353,
FP7-IDEAS-Phys.LSS 240117, FP7-PEOPLE-2007-4-3-IRGn202182.

\end{document}